\documentclass[twocolumn,a4paper,10pt]{article}

\makeatletter
\renewcommand{\@eqnnum}{[\theequation]}
\renewcommand{\@cite}[2]{({#1\if@tempswa , #2\fi})}
\renewcommand{\@biblabel}[1]{#1.\hfill}
\makeatother

\usepackage{amssymb,amsfonts,amsmath}

\usepackage{hyperref}
\usepackage{xr}
\usepackage{times}

\usepackage{graphicx}

\newcommand{\lb}[1]{\label{#1}\makebox[0pt]{\hbox to 0.5\textwidth {\fbox{#1}\hfill}} }

\newcommand{\req}[1]{Eq.\,\ref{#1}}

\newcommand{\rfig}[1]{Fig.\,\ref{#1}}

\newcommand{\e}[1]{\mathrm{e}^{#1}}
\newcommand{\I}[0]{\mathrm{i}}
\newcommand{\unit}[1]{\, \mathrm{#1}}

\usepackage{color}
\definecolor{red}{RGB}{255,0,0}

\usepackage[usenames,dvipsnames]{xcolor}

\usepackage{caption}
\begin{document}
\twocolumn[
  \begin{@twocolumnfalse}

   \begin{center} {\bf \Large Gibbs-Ringing Artifact Removal Based on Local Subvoxel-shifts } \end{center} 
   \vspace*{1cm}

{\bf
\begin{center}
\noindent 
{Elias Kellner}$^{1}$,
{Bibek Dhital}$^{1}$,
{Marco Reisert}$^1$,
\\ \vspace*{1cm}
\normalfont Corresponding author: Elias Kellner
\end{center}
}

{\small
\noindent
$^1$ {Department of Radiology, Medical Physics, University Medical Center Freiburg, Germany},\\
\\\vspace*{1cm}

}

  \end{@twocolumnfalse}
]


\section{Abstract}\vspace*{.1cm}
\bf
Gibbs-ringing is a well known artifact  which manifests itself as spurious oscillations in the vicinity of sharp image transients, e.g. at tissue boundaries.
The origin can be seen in the truncation of k-space during MRI data-acquisition.
Consequently, correction techniques like Gegenbauer reconstruction or extrapolation methods aim at recovering these missing data.
Here, we present a simple and robust method which exploits a different view on the Gibbs-phenomena.
The truncation in k-space can be interpreted as a convolution with a sinc-function in image space.
Hence, the severity of the artifacts depends on how the sinc-function is sampled.
We propose to re-interpolate the image based on local, subvoxel shifts to sample the ringing pattern at the zero-crossings of the oscillating sinc-function.
With this, the artifact can effectively and robustly be removed with a minimal amount of smoothing.

\normalfont

\noindent
Key words: {Gibbs-Ringing | Ringing-Artifact | Truncation-Artifact}

\section{Introduction}
\begin{figure}[t]
 \centering
  \includegraphics[width=\columnwidth]{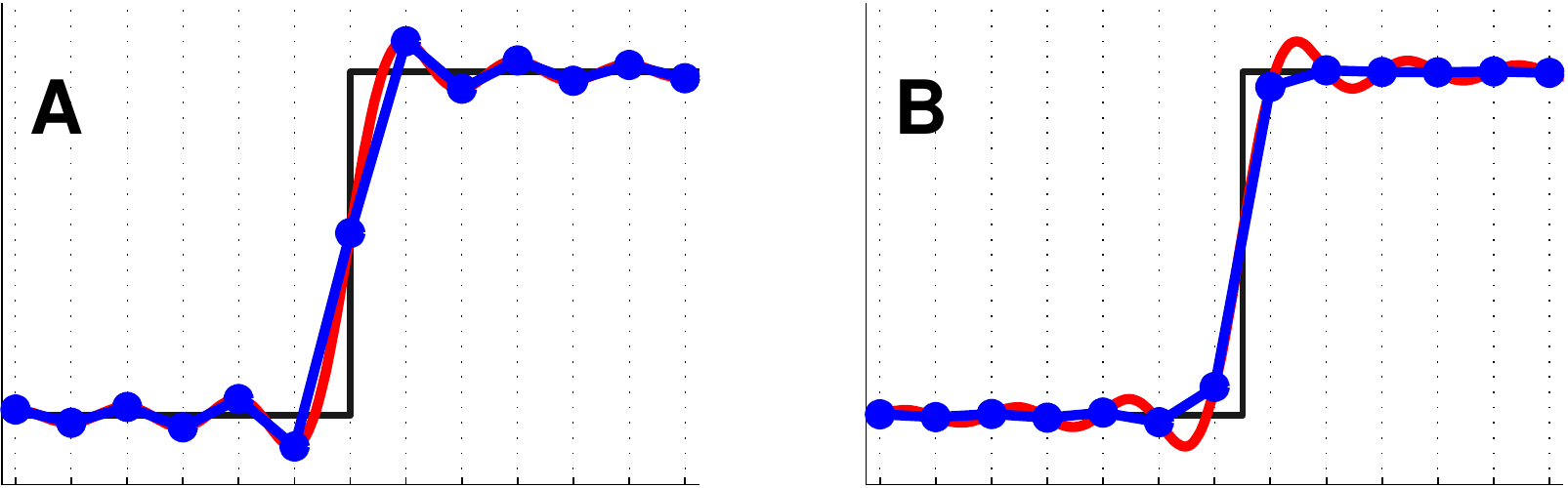}
 \caption{\label{fig::Figure_01}
An image with a single discontinuity (black edge) is reconstructed from truncated k-space data. The resulting image (blue dots) is discrete and exhibits ringing artifacts.
The amplitude of the ringing depends on whether the underlying sinc pattern arising from the windowing in k-space (red curve) is sampled at its extrema (A), or at the zero-crossings (B).
}
\end{figure}

In MRI, images are not gained directly, but with a reconstruction from acquisitions of corresponding Fourier expansion coefficients in k-space, 
\begin{equation} \label{eq::FT_image}
  I(x) = \frac{1}{N}\sum_{k=0}^{N-1} c(k) \cdot \e{\frac{-\I 2\pi k x}{N}  } \ ,
\end{equation} 
where $I(x)$ denotes the image value at voxel index $x$ and $c(k)$ are the $N$ expansion coefficients. 
In practice, only a finite number of expansion coefficients can be acquired.
This truncation of Fourier space introduces artifacts if the expansion coefficients do not decay fast enough with increasing $k$ \cite{Gottlieb1992}. 
This is the case for sharp image transitions, where all higher frequency components are required to properly reconstruct the edge.
The artifact can nicely be illustrated when reconstructing the image on a fine grid using zero-filling by setting the missing high frequency components for $k>N$ to zero.
This operation corresponds to a multiplication of the `true' k-space with a rectangular window, which is equivalent to a convolution with a sinc function in the image domain.
The side lobes of the sinc result in oscillations (`ringing') in the neighborhood of sharp edges.
The issue is illustrated in \rfig{fig::Figure_01} for the image of a single edge. 

Several approaches have been proposed to reduce ringing artifacts.
Most straightforwardly, image filters, e.g. the Lanczos $\sigma$-approximation \cite{Gottlieb1997, Jerri2000} or a median filter can be used to smooth the oscillations.
However, as filtering implies global blurring, the spatial resolution is effectively reduced, leading to a loss of fine image details.
More advanced methods have been developed based on piecewise re-reconstruction of smooth regions using Gegenbauer-Polynomials \cite{Gottlieb1997, Archibald2002}. 
A drawback of these methods is the requirement of an edge detection and potential instabilities in some applications \cite{Boyd2005} due to the involved choice of parameters.
Another modern approach consists of combined de-noising and ringing removal using using total variation constrained data exploration \cite{Block2008}.

The method we propose in this work is based on a different view on the effect: 
In fact, due to the finite number of expansion coefficients, the image is not reconstructed on a continuous domain, but on a discrete grid. 
Obviously, the strength of the ringing in the reconstructed image depends on the precise location of the edge relative to the sampling grid, i.e. how the sinc-function is sampled.
If the side lobes of the sinc-pattern are sampled at its extrema, the ringing amplitude becomes maximal, whereas it disappears when sampled at the zero crossings (see \rfig{fig::Figure_01}). 
This feature has been demonstrated in experimental data in the context of the dark rim artifact in cardiovascular imaging \cite{Ferreira2009}.
Finding the optimal subvoxel-shift for pixels in the neighborhood of sharp edges in the image can therefore minimize the oscillations. 
However, as there are multiple edges present in an image, the correction cannot be achieved by a single, global shift, but must be performed on a local basis.
In the next section, we describe a non-iterative method which conducts this task.

\section{Methods}
\begin{figure*}[ht!]
 \centering
  \includegraphics[width=2\columnwidth]{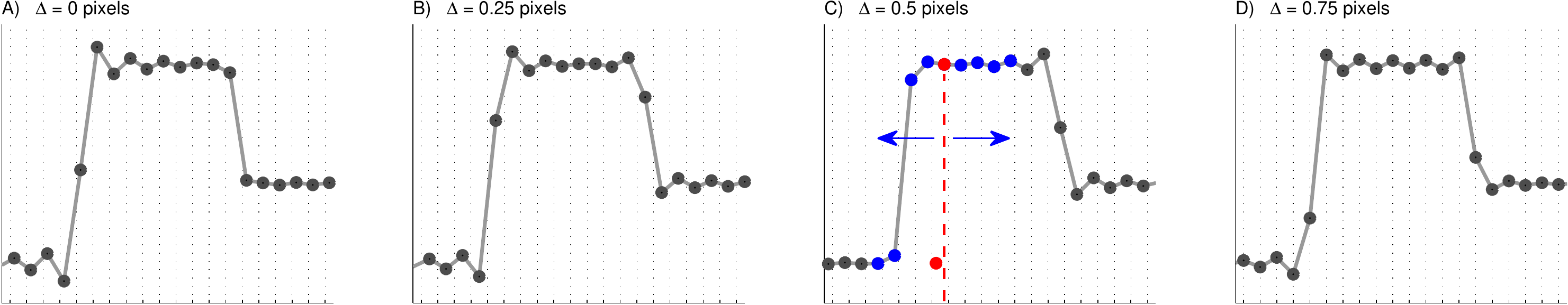}
 \caption{\label{fig::Figure_02}
1-D image with two edges. A set of images with increasing subvoxel-shifts is created.
The optimal shift is different for the two edges. For the right edge, image A) is optimal whereas it is C) for the left edge. Consequently, the optimization must be performed on a local basis, i.e. for each pixel individually.
The optimization criteria is given by the amount of ringing, which can be characterized by calculating the absolute differences in a certain range for each pixel.
In C), this is exemplary illustrated for the red pixel. The total variation is calculated separately to both, the left and right side of the center pixel. This ensures that the ringing is measured \textit{away} from the edge, and the edge itself is not contributing.
Further, the center pixel itself is excluded in order to minimize the introduced correlations between neighboring pixels.
Finally, the pixel value is recalculated by interpolating the shifted image D) at the original pixel position, which in this example is 0.5 pixels to the left side (see red dot at bottom). Mathematical details are given in the text.
}
\end{figure*}

\paragraph{One-dimensional Case}
Let $I_0(x)$ be the original, discrete image, and $c_0(k)$ its Fourier expansion coefficients.
From this, a set of $2M$ images $I_s(x)$, where ${s =  -M \dots M-1}$, with subsequent subvoxel-shifts is created by multiplication with phase-ramps in Fourier-space: 
\begin{equation} \label{eq::phase_ramp}
  I_s(x) = \frac{1}{N}\sum_{k=0}^{N-1} c_0(k) \cdot \e{\frac{-\I 2\pi}{N} k \left(x + \frac{s}{2M} \right)  }   
\end{equation} 
From this dataset, for each pixel $x$, the optimal shift which minimizes potential oscillations in the neighborhood is found. 
The corresponding measure can be calculated with any oscillation-sensitive kernel, e.g. the absolute differences in the neighborhood. 
It seems advisable to measure the ringing to both sides of the pixel individually, and select the smaller value. 
This way, in the neighborhood of an edge, the ringing is always measured in the direction opposite to the edge (see \rfig{fig::Figure_02}.
Thus, the step itself does not contribute, and interferences from closely located edges are minimized.
We decided to use a simple absolute differences approach to measure the oscillations. 
\begin{eqnarray*}
 TV^+_s(x) &=& \sum_{n=k_1}^{k_2} |I_s(x+n) - I_s(x+(n-1))| \\
 TV^-_s(x) &=& \sum_{n=k_1}^{k_2} |I_s(x-n) - I_s(x-(n-1))|
\end{eqnarray*}
where $K = [k_1,k_2]$ defines the window size which is used for measuring the amount of oscillations. 
There are reasons to exclude the central point itself, using a window of the form $K = [1,k_2]$.
This leads to more stability for points directly on the edge, and minimizes increase in noise correlation, as the oscillation measurement itself is not correlated with the 
actual point we are interested in.
The choice of the window $K$ constitutes the only parameters of the method. The results will indicate that the proposed algorithm is relatively robust against this choice.

Now, for each point $x$ the optimal shift is determined by finding the shift $s$ such that the total variation $TV$ is minimal. First, independently for the right (+)
and left (-) side:
\begin{eqnarray} \label{eq::ringMeasure}
  r^+(x) &=&  \underset{s}{\arg\min}\ TV^+_s(x) \\
  r^-(x) &=&  \underset{s}{\arg\min}\ TV^-_s(x)
\end{eqnarray} 
Finally, we decide whether the overall minimum $\text{min}(TV^+_{r^+(x)}(x),TV^-_{r^-(x)}(x))$ comes from left or right and 
this shift is determined to be the optimal shift $r(x)$. 
Now, we know the optimal shift at each image location, which minimizes TV and hence also the ringing artifact. But, finally we have
to go back to original grid, i.e. evaluating $I_{r(x)}(x')$ at the non-integer position $x' = x-r(x)/(2M)$.
In order to do so we can use any kind of alternative interpolation which is not a sinc-interpolation. Formally, we have 
\[
 I_\text{unring}(x) := I_{r(x)}(x-r(x)/(2M)) 
\]
We decided to do this final 'back interpolation' by a simple linear interpolation. In fact, one could also use alternative high-order interpolation schemes (like polynomial interpolation),
but they usually lead to ringing artifacts again. 

\paragraph{Two-dimensional Case}
The extension to the two dimensional case is not straightforward as diagonal edges produce checkerboard-like ringing patterns, as can be observed in the phantom image in \rfig{fig::Figure_phantom_edgy}).
Hence, it is not possible to find the optimal shift in both dimensions simultaneously. 
As a solution, we correct the image in both dimensions separately resulting in two one-dimensionally corrected images $J_x$ and $J_y$.
These are then combined in Fourier space to the final image $J$ via 
\begin{equation} \label{eq::2D_combination}
 J = \mathrm{FT^{-1}}   \Big\{ \mathrm{FT} \left\{J_x\right\}\cdot G_x + \mathrm{FT}\left\{J_y\right\}\cdot G_y \Big\}
\end{equation} 
where $\mathrm{FT}\{\cdot\}$ denotes the Fourier transform. 
We propose to use the `weighting functions' $G_x$ and $G_y$ with a saddle-like structure in Fourier Space of the form
\begin{equation} \label{eq::2D_combination_filers}
\begin{array}{c}
G_x = \frac{1+\cos k_y}{(1+\cos k_y) + (1+\cos k_x)} \\ \\
G_y = \frac{1+\cos k_x}{(1+\cos k_y) + (1+\cos k_x)} 
\end{array}
\end{equation} 
This way, the high frequency components along the direction of the correction are enhanced, while they are dampened along the non-corrected direction.
Due to the normalization $G_x+G_y = \bf1$, artifact-free images, where $J_x = J_y$, are left unchanged by \req{eq::2D_combination_filers}.
This ensures that minimal smoothing is introduced.

\begin{figure}[ht!]
  \includegraphics[width=.97\columnwidth]{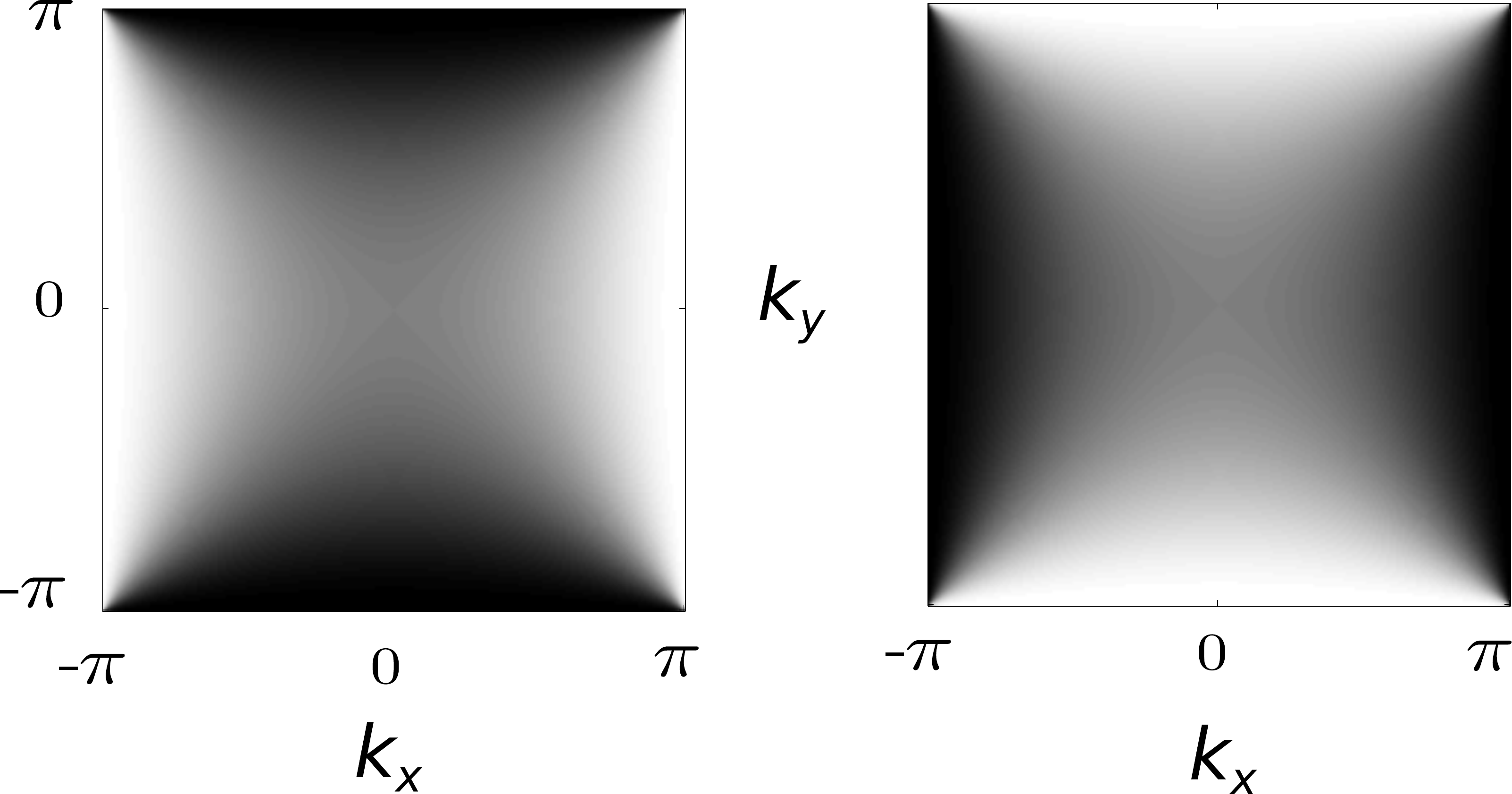}
 \caption{\label{fig::Figure_IMO}
Weighting functions $G_x$ (left) and $G_y$ (right) for multiplication with the 1D-corrected images in Fourier Space.
High frequencies are enhanced along the direction of the correction and suppressed along the other direction.
 }
\end{figure}

\paragraph{Reference methods}
We compare the proposed method against other, non-iterative filtering methods.
In order to minimize unwanted smoothing, low-pass filter with a high cutoff-frequency seem preferable. 
This can be achieved with the Lanczos sigma-approximation \cite{Gottlieb1997, Jerri2000}, where the filtered image is given by
\begin{equation} \label{eq::lanczos}
  I(x) = \frac{1}{N}\sum_{k=0}^{N-1} \left(\mathrm{sinc}\frac{k}{N}\right)^p \cdot c(k) \cdot \e{\frac{-\I 2\pi k x}{N}  } \ ,
\end{equation} 
where the parameter $p$ controls the filter strength. In this study, we use two different settings for $p$.
On one hand, we apply the standard choice of $p=1$. This induces, however, a rather strong smoothing of the image.
Hence, for a sound comparison with the proposed method, we further set $p$ such that the increase in the correlation of the noise are the same for both, the proposed method and the filtering approach. The corresponding reference noise correlation was measured in a pure-noise region on the mean-free image ${\tilde{I}(x) = I(x) - \bar{I}}$ via
\begin{equation} \label{eq::noise_corr}
  r  = \frac{ \sum_x  \tilde{I}(x)\tilde{I}(x-1 )}{\sum_x \tilde{I}^2(x)}
\end{equation} 
Another popular, non-linear filtering approach which preserves edges is given by the median filter.
For comparison, we also applied a median filter, where the `width' of the filter was fixed to a 2x2 neighborhood.

\paragraph{Numerical Phantoms}
The method was applied to two numerical phantoms (\rfig{fig::Figure_phantom_edgy}).
For the first phantom, a polygonal shape with some stripes and small structures was simulated.
Starting with a high-resolution image, the artifact was simulated by reconstructing the image from a truncated k-space with a reduction factor of 20.
Second, for a more realistic brain phantom, data from a $\mathrm{T_1}$-weighted post-contrast MRI measurement of the brain was used.
The artifact was artificially enhanced by re-reconstructing the image from a smaller k-space with a reduction factor of 4.
In both cases, a `ground truth' image without artifact, but with the same decreased spatial resolution was generated by convolving the high-resolution image with a boxcar function, and sampling the result on the corresponding low-resolution grid.
Gaussian noise with a signal-to-noise ratio of 100 was added.

\paragraph{MRI Measurements}
The method was applied to  diffusion-weighted-images (DWI) with 70 directions, $b=1000\unit{s/mm^2}$, using a gradient echo EPI sequence with $\mathrm{TE}=107\unit{ms}$, matrix size 104x104, resolution $2\unit{mm}^3$, performed on a 3T scanner (Siemens TIM TRIO, Siemens, Erlangen, Germany). 
No distortion correction was applied, as the involved correction methods already lead to significant filtering of the artifact, especially in phase direction.
Due to the different image contrast for different $b$-values, the artifact might even be amplified during post-processing of the diffusion parameters.
Therefore, we also calculated diffusion maps, without and with artifact correction.
We further applied the method to a $\mathrm{T_2}$-weighted image acquired with a turbo-spin-echo sequence, $\mathrm{TE}=109\unit{s}$, resolution 1x1x5$\unit{mm^2}$ on a 1.5T scanner (Siemens SONATA, Siemens, Erlangen, Germany).
Both dataset were acquired in the context of clinical routine, written consent was obtained to use the data for scientific use.

\section{Results}
\begin{figure*}[ht!]
 \centering
  \includegraphics[width=2\columnwidth]{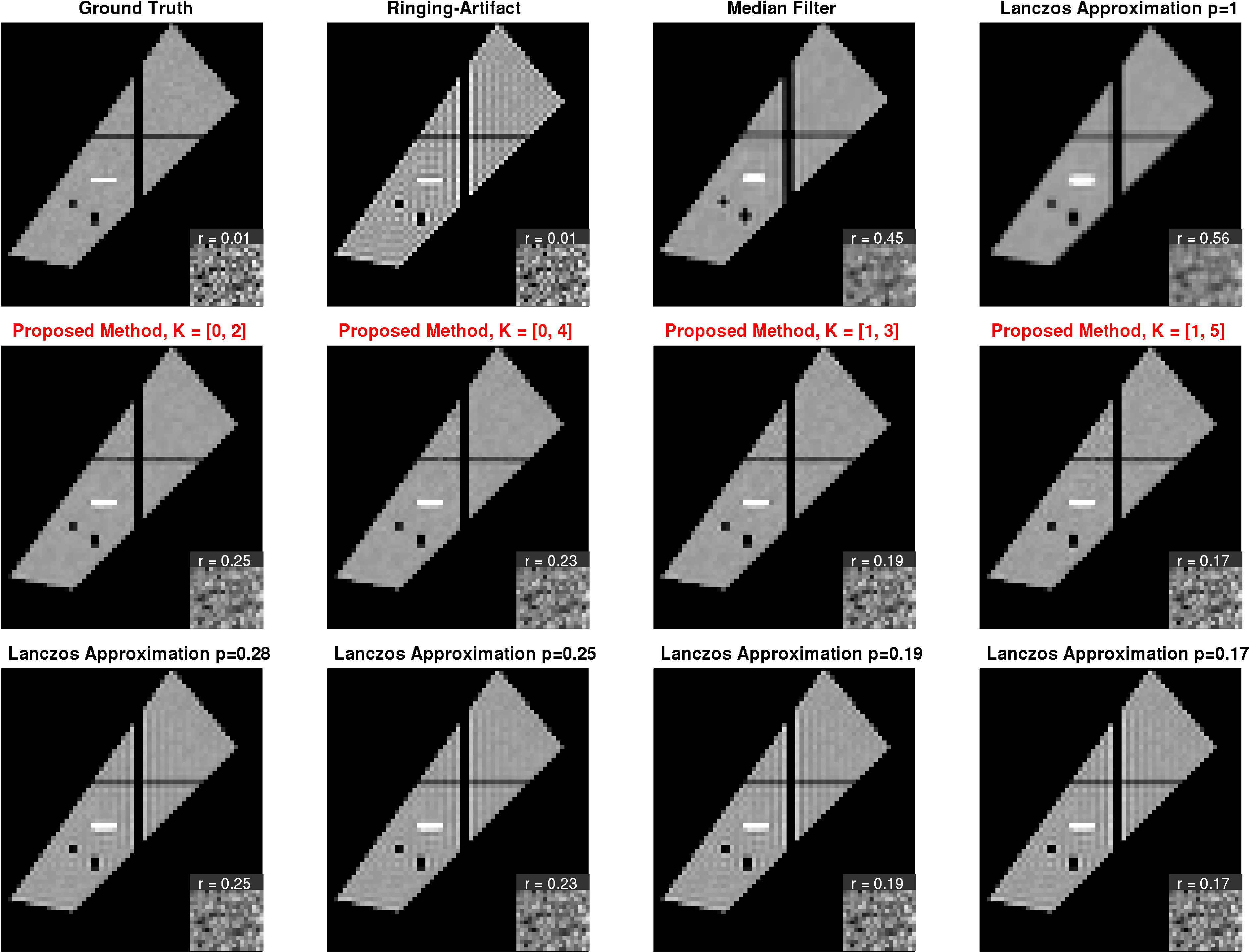}
 \caption{\label{fig::Figure_phantom_edgy}
Phantom image with multiple edges and noise. Inserts correspond to a scaled area with pure noise, correlation. 
The amount of noise correlation (given by r) reflects the strength of the smoothing introduced by the methods and their parameters.
Both, median filter and Lanczos approximation with $p=1$ lead to rather strong filtering.
The median filter additionally introduces artifacts on point-like structures (see e.g. the black dots in the images).
Results using the proposed method are given in the second row, the Lanczos approximation with corresponding equal noise correlation in the bottom row. 
With the proposed method, the artifact can virtually completely be removed with preservation of the edges and without destroying fine image details, whereas the Lanczos approximation with equal noise correlation  still suffers from significant artifacts.
}
\end{figure*}

\paragraph{Numerical Phantoms}
The results of the phantom simulations are shown in \rfig{fig::Figure_phantom_edgy}. 
We show results using the median filter, the Lanczos approximation with $p=1$, results obtained with the proposed method using different parameters, and
results using the Lanczos approximation with filter parameters $p$ adapted to yield an equal noise correlation as the proposed method.

Obviously, the median filter preserves the edges better than the Lanczos-approximation with $p=1$, at a smaller increase in the average smoothing, indicated by the smaller increase in noise correlation. However, it shows a stronger residual of the artifact, and fine image details like small, peak-like structures are destroyed.
With the proposed method, on the other hand, the artifact can effectively be removed with minimal smoothing of edges. 
The method is rather robust against the choice of the kernel parameter. 
A larger neighborhood results in less smoothing, but comes at the price of slightly reduced artifact removal.
A kernel size of $K=[1\, ,  3]$ seems to be a appropriate compromise between artifact removal and noise correlation. 
This setting is used for the application to the MRI images.

\begin{figure*}[ht!]
 \centering
  \includegraphics[width=2\columnwidth]{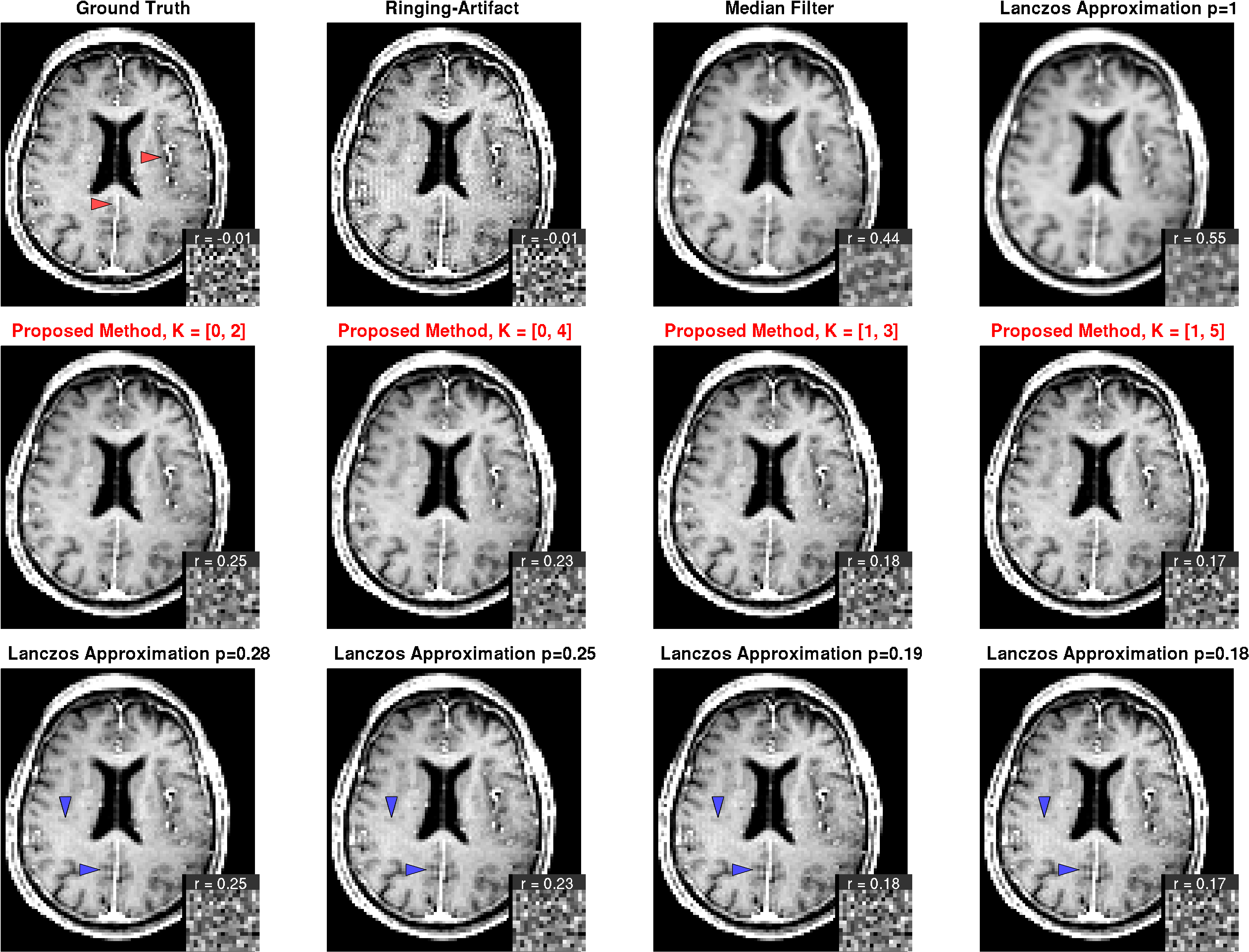}
 \caption{\label{fig::Figure_phantom_t1}
$\mathrm{T_1}$ weighted post-contrast MRI images with artificially enhanced artifact. Here, the same characteristics as in \rfig{fig::Figure_phantom_edgy} can be observed.
Both, median filter and Lanczos approximation with $p=1$ lead to blurring of image details (compare e.g. areas at red arrows in ground truth).
Again, in the Lanczos approximation with adapted parameters, there is still residual ringing visible (see e.g. blue arrows), even though the difference to the proposed method is here less pronounced compared to the phantom in \rfig{fig::Figure_phantom_edgy}.
}
\end{figure*}

These findings are basically the same for the phantom constructed from the $T_1$-weighted image in \rfig{fig::Figure_phantom_t1}. 
Also here, the artifact can most effectively be removed using the proposed method, while preserving fine image details.

\begin{figure*}[ht!]
 \centering
\large \bf \hspace*{1 cm} b=0 images \hspace*{5.5 cm} Radial diffusivity\\\vspace*{.2cm} \normalfont
  \includegraphics[width=.94\columnwidth]{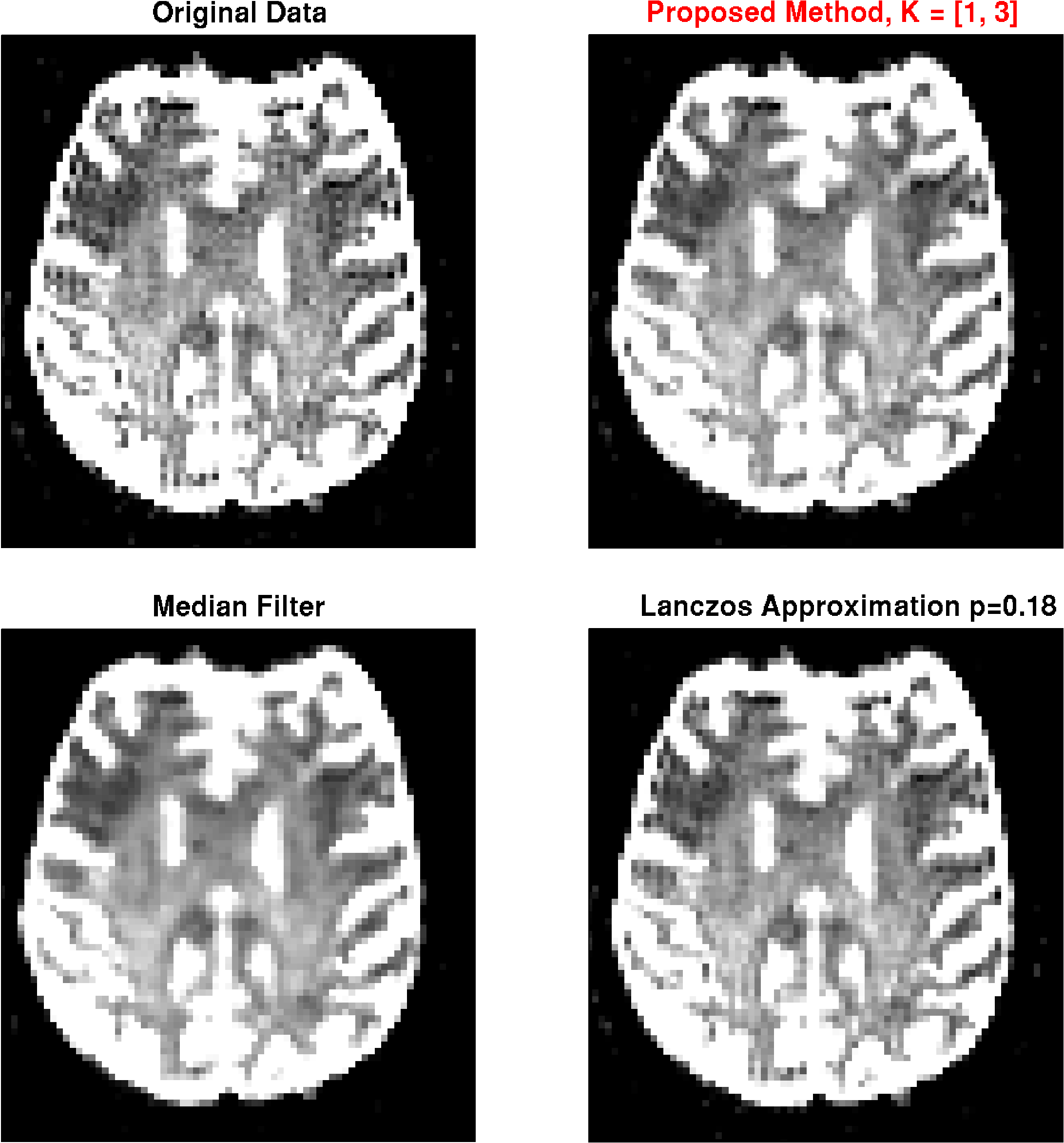}
  \hfill \vrule \hfill
  \includegraphics[width=.94\columnwidth]{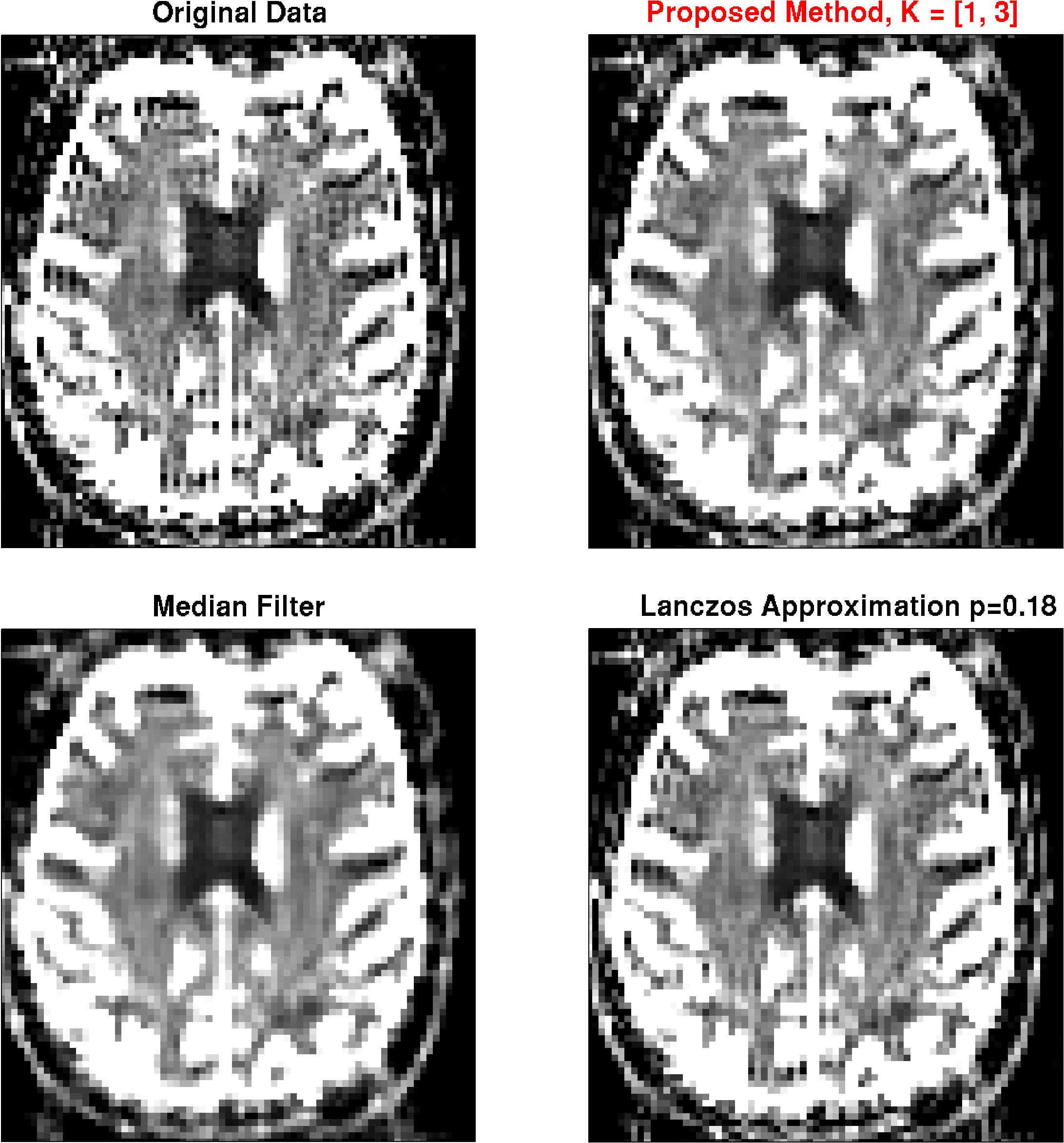}
 \caption{\label{fig::Figure_DWI_b0_and_rD_s50}
DWI measurement. The original $b=0$ image exhibits strong ringing artifacts, which are even more amplified in the calculated radial diffusivity maps (right maps).
The artifact can most effectively be removed with the proposed method at a minimal image smoothing. The Lanczos approximation shows residual artifacts.
}
\end{figure*}

\begin{figure*}[ht!]
 \centering
  \includegraphics[width=1.8\columnwidth]{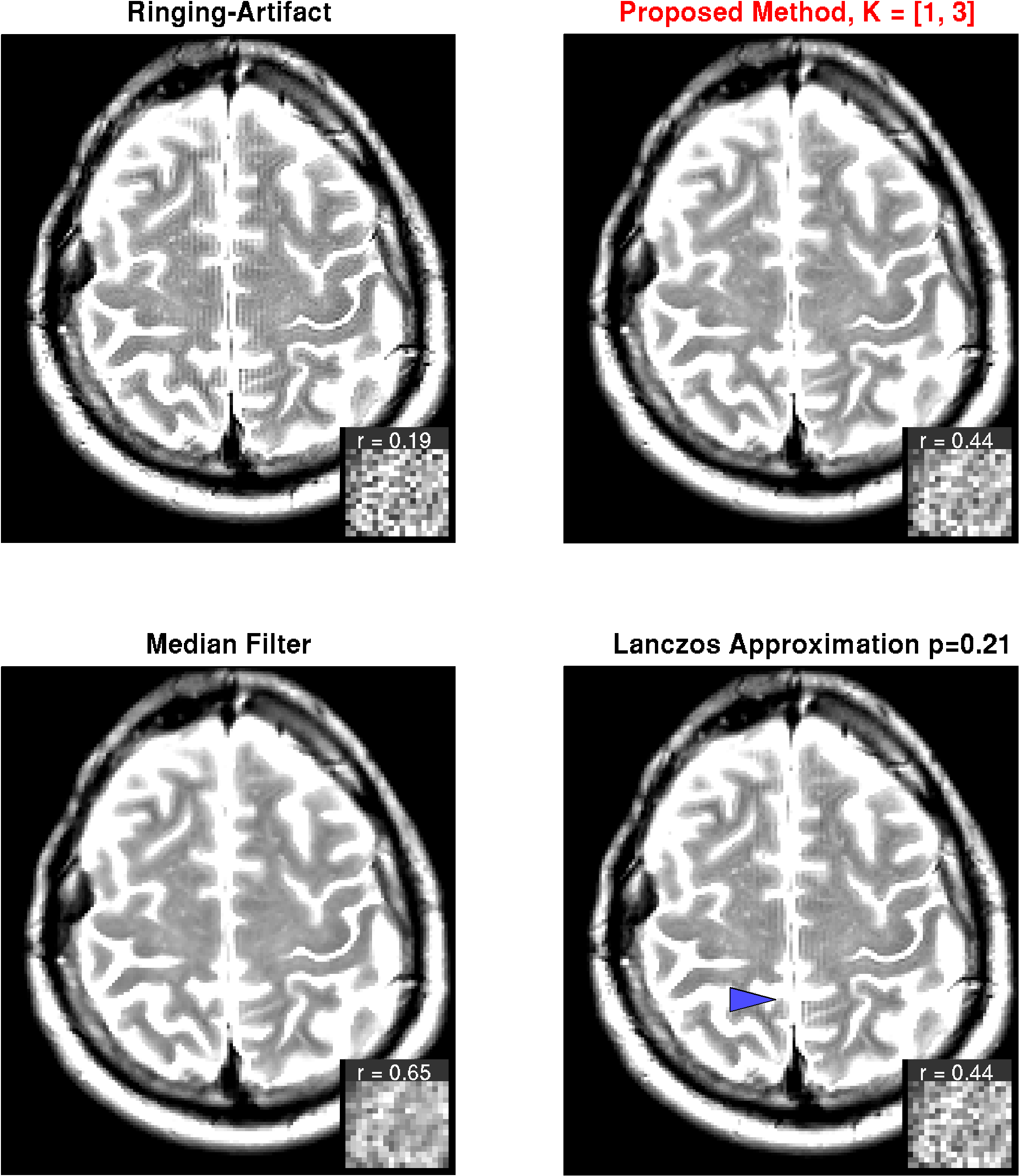}
 \caption{\label{fig::Figure_T2_01}
$\mathrm{T_2}$-weighted image with obvious ringing artifacts. With the proposed method, the artifact can effectively be removed.
The Lanczos approximation shows residual artifacts (see e.g. blue arrow).
}
\end{figure*}

\paragraph{MRI images}
The results for DWI measurements are shown for one slice in \rfig{fig::Figure_DWI_b0_and_rD_s50}. 
Apparently, the $b_0$-images exhibit strong ringing artifacts, which is even more emphasized after diffusion calculation.
The artifact can be reduced with both, the median filter and the Lanczos approximation with $p=1$, however, at the cost of strong smoothing.
With the proposed method on the other hand, the artifact can virtually completely be removed with minimal filtering.
Results from the $\mathrm{T_2}$-weighted image given in \rfig{fig::Figure_T2_01}. The findings are basically the same as for the DWI measurement.

\section{Discussion}
Even though the Gibbs-ringing artifact is omnipresent in MRI, most vendors do not include removal techniques in the standard image reconstructing pipeline to date.
One reason for this might be that standard filtering approaches inevitably reduce the effective image resolution, and more advanced methods like Gegenbauer re-reconstruction \cite{Archibald2002, Gottlieb1997, Shizgal2003, Feng2006} or data extrapolation methods \cite{Amartur1991, Constable1991} are practically difficult to handle due to their complexity and, the requirement of an edge detection, and potential instabilities induced by the dependency on the choice of the parameters.

Another approach consists of optimizations based on total variation \cite{Block2008}. These methods have proven to effectively remove the artifact, however, they treat noise and artifact equally, in contrast to the proposed method, which explicitly aims at separating both contributions. Further, the outcome of total variation applications strongly depends on the strength of the filter, which must be adapted to the respective application.

The proposed method is rather robust to the choice of its parameter, the kernel width $K$. 
Further, this parameter is independent of the image size, as oscillation pattern of the ringing occurs always in the distance of one voxel, and hence scales with the matrix size. 
The method can therefore applied to any image with a universal value for $K$.
We found that $K=[1,3]$ constitutes a good compromise between artifact removal and smoothing.

\section{Conclusions}
In this work, we presented a non-iterative method for removal of ringing artifacts based on re-sampling the image such that the source of the ringing pattern, the sinc-function, is sampled at its zero crossings. 
We demonstrated that the method effectively removes the artifact while introducing minimal smoothing.
The method has a low computational cost, consists of a rather simple mathematical framework and is very stable against the choice of its few parameters. 
This robustness suggests it as a suitable candidate for a robust implementation in the standard image processing pipeline in clinical routine.
Even though designed in the context of MRI, the method might also prove its applicability in other areas such as Fourier-based data compression algorithms.

\section{Acknowledgements}
We are grateful to Valerij G.\ Kiselev for very helpful discussions. 


\end{document}